# GRB 090423: Marking the Death of a Massive Star at z=8.2

LIN Lin[1], LIANG En Wei[2†] & ZHANG Shuang Nan[3]

[1] Department of Physics and Center for Astrophysics, Tsinghua University, Beijing 100084, China;
[2] Department of Physics, Guangxi University, Nanning 530004, China;
[3] Key Laboratory of Particle Astrophysics, Institute of High Energy Physics, Chinese Academy of Sciences, P.O. Box 918-3, Beijing 100049, China

**GRB 090423 is the new high-z record holder of Gamma-ray bursts (GRBs) with z~ 8.2. We present a detailed analysis of both the spectral and temporal features of GRB 090423 observed with Swift/BAT and Fermi/GBM. We find that the $T_{90}$ observed with BAT in the 15-150 keV band is 13.2 s, corresponding to ~ 1.4 s at z=8.2. It once again gives rise to an issue whether the progenitors of high-z GRBs are massive stars or mergers since the discovery of GRB 080913 at z=6.7. In comparison with $T_{90}$ distribution in the burst frame of current redshift-known GRB sample, we find that it is marginally grouped into the long group (Type II GRBs). The spectrum observed with both BAT and GBM is well fitted by a power-law with exponential cutoff, which yields an $E_p$=50.4±7.0 keV. The event well satisfies the Amati-relation for the Type II GRBs within their 3$\sigma$ uncertainty range. Our results indicate that this event would be produced by the death of a massive star. Based on the Amati-relation, we derive its distance modulus, which follows the Hubble diagram of the concordance cosmology model at a redshift of ~8.2.**

Received; accepted
doi:
†Corresponding author (email:)　lew@gxu.edu.cn
Supported by the National Natural Science Foundation of China under grants No. 10821061, 10733010, 10725313 and 10873002, and by the National Basic Research Program ("973" Program) of China (Grant 2009CB824800), the research foundation of Guangxi University (for EWL). This project was also in part supported by the Ministry of Education of China, Directional Research Project of the Academy of Sciences under project KJCX2-YW-T03.

Gamma-ray bursts (GRBs) are the most violent cataclysmic explosions in the universe, lasting from milliseconds to thousand seconds. The burst duration ($T_{90}$) is generally regarded as a mark of the activity duration of the central engine for this phenomenon, and the observed GRBs are classified into two groups with a division of $T_{90}$=2 seconds, i.e., long *v.s.* short GRBs [1]. It has long been speculated that long GRBs are associated with the deaths of massive stars and hence SNe ([2]-[5]). The connection between long-duration GRBs and SNe was predicted theoretically ([6] and [7]) and has been verified observationally through detecting spectroscopic features of the underlying SNe in some nearby GRBs [5]. The progenitors of short duration GRBs were thought to be mergers of two compact objects, such as NS-NS and NS-BH mergers [8-11]. The breakthrough to understand the nature of short GRBs was made with the Swift satellite, a multi-wavelength GRB mission [12]. Prompt localizations led to rapid and deep ground-based follow-up observations for a handful of short GRBs, favoring the merger models [13].

With high sensitivity of the burst alert telescope (BAT) and promptly slewing capability of the X-ray telescope (XRT) on board Swift satellite, Swift has revealed many unexpected features that challenge the conventional models [14]. The clear long-short classification scheme is also blurred with the observations by Swift/BAT and XRT. The detections of extended emission of "short" GRBs make it difficult to categorize a GRB into the "short" or a "long" group ([15-17]). It is believed that X-ray flares are of internal origin ([18-20]). The detection of late X-ray flares from short GRBs (such as 050724 [21]) also indicates that the central engines of short GRBs are not died out rapidly. With the detection of a nearby long duration GRB without SN association, GRB 060614 [22-25], which has an extended emission component up to ~100s post the BAT trigger, Zhang et al. (2007) [16] proposed a more physical classification scheme for GRBs, i.e., Type I *v.s.* Type II. The Type I GRBs are generally short with extended emission, and the Type II GRBs are generally long. However, observations of some long GRBs are also not intrinsically long. High-z GRB 080913 (*z*=6.7) is an intrinsically short, hard GRB with a duration of ~ 1 second in the burst frame [26]. Although a Type I origin invoking a BH-NS merger is not ruled out, it was suggested that GRB 080913 is more likely a Type II event at *z*=6.7 [27]. Burst duration seems to be no longer a characteristic for classification.

It is interesting that Swift detected GRB 090423 at *z*=8.2 [28], approaching the era of re-ionization of the universe. Fermi/GBM also detected this gamma-ray burst. This is the most distance object observed ever. We here investigate both the spectral and temporal features of this event observed with BAT and GBM and compare our results with typical Type I and II GRBs. The high redshift of 8.2 offers a chance to check whether the cosmology model, we also present an updated Hubble diagram for both the SNs and GRBs. Throughout, we adopt $H_0 = 70$ km/s/Mpc, $\Omega_M = 0.27, \Omega_\Lambda = 0.73$.

## 1. Data Reduction

GRB 090423 was triggered by Swift/BAT at 07:55:19 UT on the 23rd April, 2009. Swift quickly slewed to the position, with the follow up observation started at 72.5 s and 77 s after trigger for XRT and UVOT, respectively [29]. We process data from Swift/BAT and Fermi/GBM. The BAT data are downloaded from Swift official webpage. We generate 64 ms binned light curves in four energy bands (15-25 keV, 25-50 keV, 50-100 keV, and 100-150 keV). The BAT spectrum is derived by making some necessary corrections, including detectors quality, mask weighting corrections, and system error, with BAT package of Heasoft. The BAT did not slew during the observation of GRB 090423, which is convenient for spectrum subtraction. The duration of the burst is calculated by our own Bayesian Blocks method for the binned data.

GBM data are downloaded from Fermi Archive available at ftp://legacy.gsfc.nasa.gov/fermi/data/gbm/bursts/. Standard Fermi tools and Heasoft software package are used to process the data. Among twelve NaI detectors, only No.9 and 10 triggered this event. The No.9 NaI detector has larger count. Our data process is focused on this detector. We generate 0.25 s binned light curve in three energy bands (8-15 keV, 15-150 keV, and above 150 keV) from the GBM data, and calculate burst duration with our Bayesian Block method. The first 16 s of observation is used as background for $T_{90}$ calculation and spectral analysis. For seeking a better fitting result, we do the joint analysis of No.9 NaI detector and No.1 BGO detector, although there is little signal of the burst detected by BGO detectors.

## 2. Temporal properties

The 64 ms binned light curve detected with BAT and 0.25 s binned light curve detected with GBM in different energy bands are shown in Figure 1. The BAT light curves show that the emission is mostly in the energy bands lower than 100 keV. We calculate the $T_{90}$ for each band, and mark the derived $T_{90}$ in each panel, if available. The $T_{90}$ of the summed light curve in the BAT band is 13.2 seconds, consistent with that reported by Krimm et al. [29]. With a redshift of *z*=8.2, the $T_{90}$ in the burst frame is ~ 1.4 seconds, comparable to that of GRB 080913 at *z*=6.7 [26].

From BAT lightcurves we find that the $T_{90}$ of emission in the lower energy band is longer than that in the higher one. However, checking the light curve in the 8-15 keV band observed with GBM, we do not find significant detection. This fact may due to that GBM has low response to the photon in this energy band. It is bright in the 15-150 keV band, with a $T_{90}$ ~ 35 second, much longer than that measured with the BAT data. Note that after about 13 second, the emission in the GBM band is very low. With the rough background subtraction, we cannot fully confirm this extended signal is physically from the burst. Therefore, the $T_{90}$ of BAT data is adopted in the following analysis.

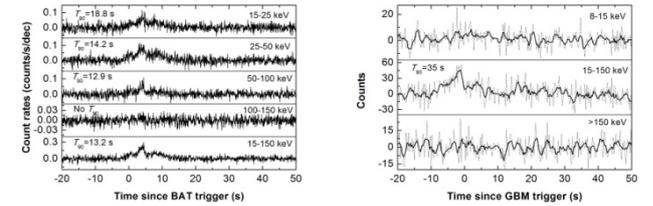

*Figure 1: left-64 ms-binned BAT light curves; right- 0.25 s-binned GBM light curves. Thick black lines are the smoothed light curves with 1s-bin.*

## 3. Spectral properties

BAT did not slew during the burst. We extract the spectrum accumulated from $T_0$-0.7 s to $T_0$+11.7 s. The spectrum is well fitted by a power-law with exponential cutoff. Our fitting results are $\alpha = 0.84^{+0.31}_{-0.34}, E_0 = 43.33^{+18.35}_{-10.74}$ keV, reduced $\chi^2 = 0.922, d.o.f. = 55$ with an $E_p$=50.38±6.98 keV, as shown in Fig. 2. The average flux in 15-150 keV is $4.84^{+6.62}_{-2.93} \times 10^{-8}$ ergs·cm$^{-2}$·s$^{-1}$. We explore the spectral evolution with the time-resolved spectrum in time intervals of [$T_0$-0.7 s, $T_0$+4.3 s], [$T_0$+4.3 s to $T_0$+8.3 s] and [$T_0$+8.3 s to $T_0$+13.3 s]. No significant spectral evolution is seen.

The joint spectrum of No.9 NaI detector and No.1 GBO detector GBM spectrum is shown in Fig. 2 (right). It has much larger uncertainty in comparison with BAT data. The fit by the same model yields $\alpha = 1.89^{+0.66}_{-0.8}$, $E_0 = 50.06^{+36.34}_{-17.75}$ keV, reduced $\chi^2 = 1.064$, $d.o.f. = 244$, marginally consistent with that derived from the BAT data with huge errors.

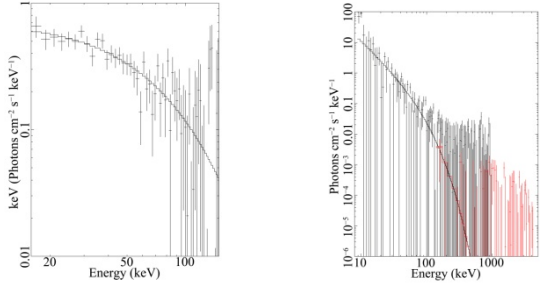

*Fig. 2: Left-BAT spectrum accumulated from $T_0$-0.7s to $T_0$+11.7 s with the best fit by an exponential cutoff power law model (line). Right- GBM spectrum by the best fit with the cutoff power law model (line).*

## 4. GRB 090423 as a Type II GRBs

It is interesting that GRB 090423 is quite similar to its partner GRB 080913 at $z$=6.7 [26]. Their $T_{90}$ are shorter than 2 sec in the rest frame, posting an issue on the classification of the two events ([27] and [30]). Note the division of long-short classification of 2 s is based on the observed $T_{90}$. We compare the distributions of observed and the intrinsic $T_{90}$ with redshift-known Swift and HETE-2 GRBs in Fig 3. Testing the bimodality of the two distribution with the algorithm by Ashman [31], we get $p<10^{-4}$ and $p$=0.004 for the $T_{90}$ distributions in the observed and the burst frames, respectively. This suggests that the bimodality for the distribution of $T_{90}$ in the burst frame is still statistically acceptable. The ratio of the probabilities of GRB 090423 in the long and short groups is 0.626: 0.374, marginally favoring the idea to classify this event into the long group (Type II).

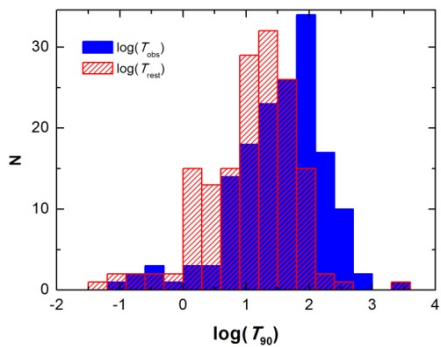

*Fig 3: $T_{90}$ distributions of Swift and HETE-2 GRBs. Swift data are from NASA official website, and HETE-2 data are from [32].*

Given $z$=8.2, the $E_{iso}$ of this event with the exponential cutoff power law model is $7.4^{+10.5}_{-4.5} \times 10^{52}$ ergs, and $E_p(1+z)$ is 463.5±64.2 keV. We examine whether GRB 090423satisfy the Amati-relation[33] with others Type II GRBs available in [27]. The result is shown in Fig. 4. It is found that GRB 090423 well satisfies the Amati-relation for the Type II GRBs within a $3\sigma$ deviation.

With the Amati-relation, we derive the distance modulus if GRB 090423 from our data. We get $\mu$=50.3$^{+1.66}_{-0.97}$. Figure 5 shows GRB 090423 in the Hubble diagram of 192 Ia supernovae [34-37] and 42 GRBs with $z$>1.4 [38]. It is clear from the figure that the Hubble diagram of the concordance cosmological model holds up to a redshift of 8, suggesting that GRBs may be promising standard candles up to re-ionization epoch[39-41].

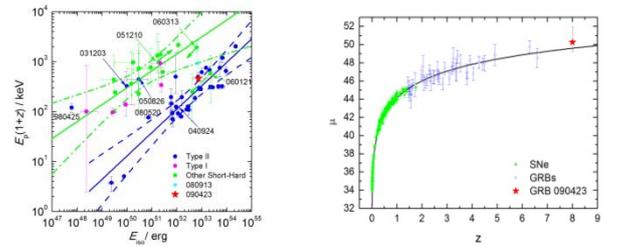

*Fig. 4: left-Comparison of GRB 090423 with the well studied Type I and Type II GRBs [47] in the $E_p(1+z)$-$E_{iso}$ plane. Blue and green solid lines are best fit result for type II and other short-hard GRBs in $3\sigma$ uncertainty range; right-Hubble diagram of 192 Ia supernovae and 42 GRBs with $z$>1.4[38]. The line is the theoretical model with $H_0 = 70$ km/s/Mpc, $\Omega_M = 0.27, \Omega_\Lambda = 0.73$*

## 5. Summary

We have investigated the spectral and temporal properties of GRB 090423 observed with Swift/BAT and Fermi/GBM and compare our results with typical Type I and II GRBs. We find that the $T_{90}$ of its emission in the lower energy band is longer than that in the higher one. The summed light curve in the BAT band (15-150 keV) is 13.2 s. With a redshift of $z$=8.2, the $T_{90}$ in the burst frame is ~ 1.5 seconds, comparable to that of GRB 080913 at $z$=6.7 [26]. The GBM data show that it is bright in the 15-150 keV band, with an observed $T_{90}$ of ~ 35 s, much longer than that measured with the BAT data. Note that after about 13 second, the emission in the GBM band is very low. With the rough background subtraction and poor response to low energy photons of GBM as seen in the lightcurves and in the observed spectrum, we cannot fully confirm this extended signal is physically from the burst.

Testing the $T_{90}$ distribution in the burst frame for the current GRB sample with redshift measurement, the bimodality exists, and GRB 090423 is marginally grouped into the long group (Type II GRBs).

The spectra of GRB 090423 observed with BAT and GBM are well fitted by a power-law with exponential cu-toff. It well satisfies the Amati-relation for the Type II GRBs within a $3\sigma$ deviation. Based on the Amati-relation, we derive the distance modulus of GRB 090423, and show that the Hubble diagram of the concordance cosmology model holds up to a redshift of ~8.

## Acknowledge

We acknowledge Zhang Bing and Zhang Binbin for helpful discussion. We also appreciate helpful suggestions of the referees.